\documentclass[journal,comsoc]{IEEEtran}

\title{\acrshort{sdn} enabled \gls{icn} as a Service prefetching mechanism for \gls{http} based services. The scalable video streaming case}
\author{
  \IEEEauthorblockN{Jordi Ortiz, Antonio Skarmeta}
\\
  \IEEEauthorblockA{jordi.ortiz@um.es, skarmeta@um.es}
}

\usepackage{graphicx}
\usepackage{url}

\usepackage{listings}

\usepackage[numbers]{natbib}

\usepackage{glossaries}
\makeglossaries
\newacronym{ahs}{AHS}{Adaptive HTTP Streaming}
\newacronym{avc}{H.264/AVC}{Advanced Video Coding}
\newacronym{arp}{ARP}{Address Resolution Protocol}
\newacronym{api}{API}{Application Programming Interface}
\newacronym{as}{AS}{Autonomous System}
\newacronym{bgp}{BGP}{Border Gateway Protocol}
\newacronym{bod}{BoD}{Bandwidth On Demand}
\newacronym{bps}{bps}{Bits per Second}
\newacronym{ccn}{CCN}{Content-Centric Networking}
\newacronym{cdn}{CDN}{Content Delivery Network}
\newacronym{cdnaas}{CDNaaS}{Content Delivery Network as a Service}
\newacronym{cif}{CIF}{Common Intermediate Format}
\newacronym{cmt}{CMT-SCTP}{Concurrent MultiPath Transfer for Stream Control Transmission Protocol}
\newacronym{coap}{CoAP}{Constrained Application Protocol}
\newacronym{cpu}{CPU}{Central Processing Unit}
\newacronym{dash}{DASH}{Dynamic Adaptive Streaming over HTTP}
\newacronym{dht}{DHT}{Distributed Hash Tables}
\newacronym{dmca}{DMCA}{Digital Millenium Copyright Act}
\newacronym{dos}{DoS}{Denial of Service}
\newacronym{dns}{DNS}{Domain Name System}
\newacronym{dona}{DONA}{Data-Oriented Network Architecture}
\newacronym{dpid}{dpid}{OpenFlow Datapath ID}
\newacronym{drm}{DRM}{Digital Right Management}
\newacronym{ebgp}{eBGP}{External Border Gateway Protocol}
\newacronym{eid}{EID}{Endpoint Identifier}
\newacronym{erdf}{ERDF}{European Regional Development Fund}
\newacronym{ethane}{Ethane}{Ethane: A Protection Architecture for Enterprise Networks}
\newacronym{fi}{FI}{Future Internet}
\newacronym{forces}{ForCES}{Forwarding and Control Element Separation}
\newacronym{fpa}{FPA}{Framework Partnership Agreement}
\newacronym{fps}{FPS}{Frames per second}
\newacronym{gsmp}{GSMP}{General Management Switch Protocol}
\newacronym{hd}{HD}{High Definition}
\newacronym{hdht}{H-DHT}{Hierarchical Distributed Hash Tables}
\newacronym{hevc}{HEVC}{High Efficiency Video Coding}
\newacronym{himalis}{HIMALIS}{Heterogeneity Inclusion and Mobility Adaptation through Locator ID Separation}
\newacronym{hip}{HIP}{Host Identity Protocol}
\newacronym{hls}{HLS}{HTTP Live Streaming}
\newacronym{http}{HTTP}{HyperText Transfer Protocol}
\newacronym{https}{HTTPS}{HTTP over TLS}
\newacronym{ibgp}{iBGP}{Internal Border Gateway Protocol}
\newacronym{icmp}{ICMP}{Internet Control Message Protocol}
\newacronym{icn}{ICN}{Information Centric Networking}
\newacronym{icnaas}{ICNaaS}{Information Centric Network as a Service}
\newacronym{iec}{IEC}{International Electrotechnical Commission}
\newacronym{ietf}{IETF}{Internet Engineering Task Force}
\newacronym{iot}{IoT}{Internet of Things}
\newacronym{ip}{IP}{Internet Protocol}
\newacronym{irtf}{IRTF}{Internet Research Task Force}
\newacronym{isdn}{ISDN}{Integrated Services Digital Network}
\newacronym{iso}{ISO}{Organization for Standardization}
\newacronym{isp}{ISP}{Internet Service Provider}
\newacronym{itu}{ITU-T}{ITU’s Telecommunication Standardization Sector}
\newacronym{kbps}{kbps}{kilobit per second}
\newacronym{libav}{LibAV}{Open source audio and video processing tools}
\newacronym{lisp}{LISP}{Location/ID Separation Protocol}
\newacronym{lru}{LRU}{Least Recently Used}
\newacronym{lxc}{LXC}{Linux Containers}
\newacronym{mac}{mac}{Media Access Control}
\newacronym{mane}{MANE}{Media Aware Network Element}
\newacronym{mano}{MANO}{Management and Orchestration}
\newacronym{mec}{MEC}{Mobile Edge Computing}
\newacronym{mofi}{MOFI}{Mobile Oriented Future Internet}
\newacronym{mpd}{MPD}{Media Presentation Description}
\newacronym{mpeg}{MPEG}{Moving Picture Experts Group}
\newacronym{mtu}{MTU}{Maximum Transmission Unit}
\newacronym{nalu}{NALU}{Network Abstraction Layer Unit}
\newacronym{nat}{NAT}{Network Address Translation}
\newacronym{netconf}{NetConf}{Network Configuration Protocol}
\newacronym{nepi}{NEPI}{Network Experimentation Programming Interface}
\newacronym{netinf}{NetInf}{Network of Information}
\newacronym{ndn}{NDN}{Named Data Networking}
\newacronym{nfv}{NFV}{Network Function Virtualisation}
\newacronym{nict}{NICT}{National Institute of Information and Communication Technology of Japan}
\newacronym{noc}{NOC}{Network Operations Center}
\newacronym{nos}{NOS}{Network Operating System}
\newacronym{nodeid}{NodeID}{Node Identity Internetworking Architecture}
\newacronym{nren}{NREN}{National Research and Education Network}
\newacronym{ntp}{NTP}{Network Time Protocol}
\newacronym{oidc}{OIDC}{OpenID Connect}
\newacronym{onos}{ONOS}{Open Network Operating System}
\newacronym{opencache}{OpenCache}{OpenCache}
\newacronym{openflow}{OpenFlow}{OpenFlow}
\newacronym{opensig}{OPENSIG}{Open Signalling Working Group}
\newacronym{osi}{OSI}{Open Systems Interconnection}
\newacronym{ott}{OTT}{Over-the-top}
\newacronym{ovs}{OVS}{Open vSwitch}

\newacronym{pf}{SCTP-PF}{Stream Control Transmission Protocol Potentially Failed}
\newacronym{ple}{PLE}{PlanetLab Europe}
\newacronym{psirp}{PSIRP}{Publish-Subscribe Internet Routing Paradigm}
\newacronym{psnr}{PSNR}{Peak Signal Noise Ratio}
\newacronym{pursuit}{PURSUIT}{Publish Subscribe Internet Technology}
\newacronym{qoe}{QoE}{Quality of Experience}
\newacronym{qos}{QoS}{Quality of Service}
\newacronym{qp}{QP}{quantization parameter}
\newacronym{rangi}{RANGI}{Routing Architecture for the Next Generation Internet}
\newacronym{rcp}{RCP}{Routing Control Platform}
\newacronym{rda}{RDA}{Rate Determination Algorithm}
\newacronym{rest}{REST}{REpresentation State Transfer}
\newacronym{rloc}{RLOC}{Routing Locator}
\newacronym{rpc}{RPC}{Remote Procedure Call}
\newacronym{rpl}{RPL}{Routing Protocol for Low Power and Lossy Networks}
\newacronym{rtmp}{RTMP}{Real-time Messaging Protocol}
\newacronym{rtp}{RTP}{Real-time Transport Protocol}
\newacronym{rtsp}{RTSP}{Real-time Streaming Protocol}
\newacronym{rtt}{RTT}{Round Trip Time}
\newacronym{sap}{SAP}{Session Announcement Protocol}
\newacronym{sctp}{SCTP}{Stream Control Transmission Protocol}
\newacronym{sdn}{SDN}{Software Defined Networking}
\newacronym{sdp}{SDP}{Session Description Protocol}
\newacronym{shvc}{SHVC}{Scalable High-Efficiency Video Coding}
\newacronym{sla}{SLA}{Service Level Agreement}
\newacronym{smartfire}{smartFIRE}{Enabling SDN ExperiMentAtion in WiReless Testbeds exploiting Future Internet Infrastructure in South KoRea and Europe}
\newacronym{svc}{H.264/SVC}{Scalable Video Coding}
\newacronym{tcp}{TCP}{Transmission Control Protocol}
\newacronym{udp}{UDP}{User Datagram Protocol}
\newacronym{uri}{URI}{Uniform Resource Identifier}
\newacronym{url}{URL}{Uniform Resource Locator}
\newacronym{vbr}{VBR}{Variable Bit-Rate}
\newacronym{vceg}{VCEG}{Video Coding Experts Group}
\newacronym{vcr}{VCR}{Video Cassette Recorder}
\newacronym{vlan}{VLAN}{Virtual Local Area Network}
\newacronym{vod}{VoD}{Video on Demand}
\newacronym{webrtc}{WebRTC}{Web Real-Time Communications}
\newacronym{wlan}{WLAN}{Wireless Local Area Network}
\newacronym{www}{WWW}{World Wide Web}
\newacronym{xml}{XML}{eXtensible Markup Language}

\begin{document}

\maketitle

\begin{abstract}
The importance of \acrshort{http} in today's networks is undisputed.
As a solution to enhance \acrshort{qos} and enhance scalability \acrshort{cdn} networks have been designed and deployed. Recently, a new paradigm known as \gls{icn} has been envisioned focusing the network routing on the content itself instead of the geographical attachment of addresses.
\glspl{sdn} have been researched for the last 10 years as enablers of \gls{fi} architectures in general and of \acrshort{icn} in particular. We have already proposed the \gls{icnaas} architecture to provide with end-to-end \gls{http} \gls{icn} alike transmission with \gls{http} in-network caching which has been thoroughly evaluated in this paper.
This paper also proposes a nouveau mechanism, which we have named prefetching mechanism, to enhance data transmission rates for first requesters that can usually not benefit from previous access to the same content. 
To evaluate and demonstrate the possibilities offered by the proposal \acrshort{svc} video streaming with \acrshort{dash} has been employed.

\end{abstract}


\section{Introduction}
\gls{ip} has become the de facto standard for network level communications as a consequence of the \textit{Internet} expansion. The growth of the \textit{Web} lead to the adoption of \gls{http} as the application level standard, therefore adopting \gls{tcp} at the transport level. Multimedia and in general bandwidth hungry or delay constrained applications have historically employed datagram based communications to avoid the limitations imposed by reliable connections. Despite the inadequacy of \gls{tcp}, the increase in available bandwidth and reduction in \gls{rtt} have allowed its expansion. It is considered that a connection with twice available bandwidth in relation to the bandwidth needed can be entrusted to \gls{tcp} without noticeable drawbacks\cite{Bing2008}. 

As a mean to enhance \gls{http} performance and take advantage of the geographical \gls{ip} address range distribution, the \gls{cdn} services appeared as caching services near the network edge, hence enhancing the perceived \gls{qos} by replicating the content into network edges' caches while ensuring content fairness. The scalability of the \textit{Internet}, and in particular of multimedia and video content, is tightly attached to these kind of deployments.

Recently the \gls{fi} has seen proposals that among others, try to break the ossification of \gls{ip} networks and therefore of the \textit{Internet}. Among those, the \gls{icn}\cite{Brito2013} breaks with the geographical implication of the network address by leveraging routing on the content instead of the location so focusing on what and not on where. In general, these systems provide with in-network caching and are primarily represented by the \gls{ccn}\cite{jacobson2009nnc}. Also as part of the \gls{fi}, the \gls{sdn} paradigm focuses on providing with a programmable network therefore facilitating the adoption of new protocols and architectures such as the \gls{icn}. In addition, \gls{sdn}\cite{Astuto2014} changes the usual distributed network control, managed by network elements speaking through control protocols, for a centralized control system in which the network elements become simple executioners of the orders received via a control protocol, usually \gls{openflow}\cite{Pfaff2011Openflow11}.

On the other hand video streaming services based on well known \gls{rtp}, \gls{rtsp} and \gls{udp} combination have been gradually cast out in favour of \gls{tcp} based connections like the ones in \gls{rtmp} and lately the \gls{http} based plethora of video streaming protocols. The later have clearly become more and more popular these days partly because their simplicity and their capability to transparently take profit of all the enhancements made to \gls{http} after twenty years. In particular, the \gls{http} based video transmission techniques take profit of \glspl{cdn} and \gls{http}'s capabilities to traverse proxies and firewalls. The \gls{http} based video streaming is a killer application for a \gls{cdn} alike system and therefore should be part of any proposal.

In general most of the \gls{fi} approaches leverage on a clean-slate solution that make them difficult to be straightforward adopted. On \cite{krishna2018user} an \gls{icnaas} architecture is presented migrating the well known and vastly adopted \gls{cdn} concept to a higher level by offering an 'as a Service' content centric way of defining new \gls{http} caching systems by means of an \gls{sdn} application, hence leveraging on \gls{sdn} to steer the traffic to the destination based on the requested \gls{url}. Unlike other proposals, the \gls{icnaas} does not translate the \gls{http} traffic to an intermediate representation but adopts a half-way approach between the clean-slate and the legacy by employing a proxy in order to inspect the \gls{url} and inform the \gls{sdn} controller about the intention to download a certain content. The caching systems employed in this solution are \gls{http} based and therefore accept any existing software cache, appliance or even the already deployed \glspl{cdn}.

In this paper we offer an extension to the aforementioned system in which the centralized control of the network makes use not only of its topology knowledge but also of the metadata of the requested content to prefetch information from the source content provider, hence potentially increasing the system's cache hit ratio.

The remainder of this paper is organised as follows. Section \ref{sec:soa} introduces basic concepts needed to understand the rest of the publication as well as state of the art of what other researchers have proposed. Then Section \ref{sec:icnaas} introduces the \gls{icnaas} architecture and introduces its internals on top of which the prefetching proposal is designed and evaluated. Sections \ref{sec:prefetch} and \ref{sec:results} present the prefetching mechanism for \gls{svc} on top of \gls{dash} and the results of our proof of concept for such a mechanism respectively. Finally Section \ref{sec:conclusions} provides with conclusions and introduces future work lines.

\section{State of the Art}
\label{sec:soa}


'A \gls{cdn} is a collection of network elements arranged for more effective delivery of content to end-users'\cite{Pathan2006}. \glspl{cdn} are usually geographically distributed caching systems that are offered to content providers and \glspl{isp} to move their contents closer to the final user, hence \glspl{cdn} are usually third parties\cite{Pathan2006}. 



\glspl{cdn} are traditionally implemented through a mix of techniques like HTTP redirection, DNS load distribution, anycast routing, and application-specific solutions, among others. As a result, a complex distributed system is in charge of redirecting users' requests to clusters of network caches. The decision on which cache should receive the content request is usually based on the communication endpoints regardless the content being requested.

The \gls{cdn} approach which emerged as an effect of Internet's evolution is also constrained by the protocols on which it is based. As an alternative to the 'ossified' \gls{ip} end-to-end and geographically attached communication system, the \gls{icn} paradigm advocates for delivering requested resources based on themselves and independently from the data transport~\cite{ietf2012irtf}. This can potentially increase the efficiency and scalability of content distribution, but it typically requires the deployment of state-of-the-art protocols like CCNx~\cite{jacobson2009nnc}.

The \gls{icn} approach places information pieces (content) as the central element of the network, making clients declare their “interest” on content pieces and providers to offer and deliver them to the intermediate network elements which, in turn, collaborate to deliver the requested content pieces to those clients, therefore advocating for what is known as \textit{in-network} caching.


Among the vast bibliography related to \gls{icn} two are the most related contributions to the work that is going to be introduced in the following Section \ref{sec:icnaas}. Both studies employ \gls{dash} as a mean for \gls{vod}.

Authors in \cite{Georgopoulos2014opencache} introduce the 'Cache as a Service' concept and based their OpenCache also in \gls{sdn} for \gls{vod}. In their proposal the control and decision of what content is to be cached is delegated to the \gls{isp} \gls{noc} that has the knowledge and ability to optimize network utilization and possibly save its precious up-link bandwidth to other \gls{isp}. The authors also introduce the concept that given certain \glspl{sla}, the solution could be also exposed to content providers such as \glspl{cdn}. We actually agree with this view and even go further thinking that what should be offered to content providers is the instantiation of exclusive \gls{icn}/\gls{cdn} instances that can be customized and controlled by the providers and implemented in the \gls{isp} premises, who could influence the behaviour of the hosted \glspl{icn} by means of the offered caching algorithms.

Similarly, authors in \cite{Grandl2013} evaluate the \gls{vod} paradigm on \gls{fi} architectures, in particular \gls{ccn} and leveraging on \gls{sdn}. The motivation for their proposal and the previous study is the collision of interests between client based rate adaptation in \gls{dash} and the in network transparent content caching paradigm and how the later confuses the former. In the study made by the authors, they stated that even in a stable environment the video representation selected by the \gls{dash} client was not stable, obtaining an oscillating pattern. They also conclude that \gls{dash} chunk requests are spread among the available rates unlikely repeating them on the retrieval of the same video any time in the future.

In their evaluation of \gls{ccn} for \gls{dash} the authors introduce a proxy to translate the \gls{http} requests to \gls{ccn} interests and employ a \gls{sdn} controller for network traffic. The findings are that ICN has two negative impacts on \gls{dash}: reduced cache hit rate and imprecise rate estimation in the client due to the difference appreciated in channels to cache and to \gls{vod} server. In their solution proposal, the authors rewrite the \gls{mpd} offered to the client extending it with the information related to the cache. Similarly our proposal in \cite{krishna2018user} and summarized in section \ref{sec:icnaas} employs a proxy to feed the \gls{sdn} controller and in particular the \gls{icnaas} application with the \gls{url} to be able to steer the traffic to the appropriate endpoint but without the burden of transport stream to datagram translation.

In a non-clean slate approach, authors in \cite{Fayazbakhsh2013} propose an \gls{icn} architecture with its focus on similar premises of those of \gls{icnaas}, providing with a non-clean slate \gls{icn} system in this case completely based on already existing and employed \gls{cdn} technologies and techniques, the incrementally deployable \gls{icn} or idICN. In this case the content must be explicitly registered into a proxy which in turn registers it in \gls{dns}, in addition, a system to auto-configure (such as proxy auto-config PAC) the clients so that the border proxy is reached is needed. There is also an agreement with the authors on the importance on coordinating network traffic engineering, and in particular that of \glspl{isp} , with content engineering to fulfill the goals that each point of view have.

Authors in \cite{Pathan2006} highlighted some future directions that are directly related to the \gls{icnaas} architecture: Service composition highly motivated by user preferences; Dynamic Content; and An adaptive CDN for media streaming.

Nowadays content delivery, with emphasis in video streaming, is the major source of bandwidth consumption in the Internet, so efficient and effective content distribution is a key aspect to deploy bandwidth demanding services at large scales. Cisco\cite{CiscoMobile2017} estimates that 
'\textit{With the emergence of popular video-streaming services that deliver Internet video to the TV and other device endpoints, CDNs have prevailed as a dominant method to deliver such content. Globally, 70 percent of all Internet traffic will cross CDNs by 2021, up from 52 percent in 2016. Globally, 77 percent of all Internet video traffic will cross CDNs by 2021, up from 67 percent in 2016}'.

\gls{dash}\cite{iso2014dash} is \gls{mpeg}'s standardized approach to \gls{http} based video streaming techniques, one of its more important characteristics is that it is codec agnostic which allows it to evolve embracing any codec evolution present and future, only the compatibility with the ISO Base Media File Format for media storage is mandatory. It reuses already existing and consolidated technologies, such as \gls{http} and \gls{xml}, to enable efficient and high-quality media delivery through networks. The idea behind \gls{dash} is to create redundant metadata, which provides extra functionality with insignificant overload to the network architecture and service provider. Thus, it delegates to the client most of the complexity.

To ease the streaming process \gls{dash} might split the media into small chunks of data which are indexed with a so called \gls{mpd} file. A single \gls{mpd} file is able to contain different representations for the same content with different characteristics. Therefore, a client can easily select or switch between different versions of the delivered (streamed) file, with different qualities like bit-rate or picture size. 


That said, the main advantage of \gls{dash} over other (non \gls{http} based) existing streaming mechanisms is that both, the chunks and \gls{mpd} files can be easily stored in already existing \gls{http} caching infrastructures. In addition, \gls{dash} will straightforward take advantage of almost any optimisation that could have been applied to the existing infrastructures such as the \glspl{cdn}. For this and other applications, the \gls{dash} standard defines profiles and allows different modes (live, on-demand, and others) providing with interoperability and suitability for different services.

\gls{svc} is the scalable extension to \gls{avc} and among its characteristics, its layered architecture with dependencies between layers and the backward compatibility of its base layer are highlighted. Although \gls{svc} was not widely adopted by the industry, there are movements to provide with new generation of scalable codecs such as \gls{shvc}.

It is clear the interest on integrating \gls{dash} video streaming in \gls{icn} architecture such as \gls{ccn}. We argue nevertheless that a end-to-end \gls{http} \gls{icn} capable architecture is needed and that can be easily and optimally implemented by employing \gls{sdn}. Thanks to that system, mechanisms like the one presented in this paper for prefetching mechanisms aware of the type of content being retrieved can be implemented.



\section{\gls{icnaas}}
\label{sec:icnaas}
This section introduces the \gls{icnaas} architecture presented in \cite{krishna2018user} in which the later proposed prefetching mechanism and presented in Section \ref{sec:prefetch} is integrated.

\subsection{Motivation}
\glspl{cdn} are already coping with the challenge of scaling \gls{http} based services as well as multimedia streaming services in a non-disruptive approach. Nevertheless, \glspl{cdn} rely on techniques which have been appearing and evolving as patches to initial design limitations, such as \gls{http} redirection or \gls{dns} load distribution, and the ossification of the \gls{ip} environment. On the business perspective, \glspl{cdn} are usually provided by third parties or only available to big companies resourceful enough. The possibility to instantiate on-demand \gls{cdn} alike mechanisms as a service by any content provider is a key characteristic to break the distance between smaller and bigger content providers, this approach could even allow with the means for content delivery optimization for internal use like intranets.


The Internet has become a reliable and mandatory service that stands on top of a patched and non-reliable mechanism. Although the temptation to clean-slate the system and produce new paradigms and architectures is very big, the solutions to today's problems as well as the enhancement proposals, must be backward compatible and realistic.


The \gls{icn} paradigm advocates for delivering requested resources based on their name and independently from the data transport~\cite{ietf2012irtf}.

On the other hand, in the last few years we have witnessed the rise of \gls{sdn} and the high momentum that has gained~\cite{nunes2014survey}. By means of a logically centralized controller that maintains the global view of the network and exposes a programmatic interface, \gls{sdn} offers huge opportunities for network programmability, service automation, and simplified management.

With all of that in mind, we proposed an architecture to deploy \gls{icn} as a service to be provided by \gls{sdn} enabled networks while being completely backward compatible with the legacy \gls{http} end-to-end approach.

In order to steer \gls{http} traffic based on its \gls{url}, the \gls{url} itself needs to be known. From the \gls{tcp} connection perspective, the \gls{url} is not known until the 3-way \gls{tcp} handshake has been accomplished. The \gls{tcp} splicing \cite{TcpSplicingBinder2015} or delayed binding\cite{DelayedBindingKopparapu2002} is a technique widely used and introduced by proxies to leverage on the kernel the rest of the communication once a milestone has been reached, reducing resource consumption. Similarly, in the \gls{icnaas} the controller delegates on the proxy the initiation of the \gls{tcp} connection and the inspection of the \gls{url} inside the \gls{http} request by redirecting any client connection to its nearest proxy. At that point the Controller is informed by the proxy (employing message 6 in Table \ref{tab:Rest}) and can select the cache and prepare the connection from the proxy to the cache for that precise request, the proxy is informed about the matching rules for that connection and then it can connect to the cache. The decision is done based on the \gls{icn} instances running, as well as the \gls{url} being requested. 

To make the process transparent to both ends, \gls{ip} and \gls{mac} rewriting capabilities exposed by the \gls{sdn} are employed so that the connection from the controller to whatever \gls{ip} and \gls{tcp} port are rewritten to cache's \gls{ip}, \gls{tcp} port and \gls{mac}, thus the cache's operating system transparently accepts the incoming packets. The details about the design and solution are presented below.


The \gls{icnaas} architecture is designed to be content independent with the only restriction set to use the \gls{http} protocol which in turn was the main objective to be achieved. As introduced in previous section, video streaming is expected to be the highest bandwidth consuming content type in the near future and the adoption of \gls{http} video streaming is a fact already, therefore, approaching the \gls{vod} as first use case seems reasonable, moreover taking profit of the correlation between the \gls{mpd} downloaded as a bootstrapping of the video streaming process and the highly probable correlation with the video chunks to be downloaded, allows the definition of advanced caching policies and advanced techniques like prefetching.

We consider that customized \gls{cdn} creation is a service to be provided as part of \gls{isp} services and inside their premises. Benefits for \gls{isp} go from uplink bandwidth consumption reduction to third parties (\glspl{cdn}) and diversification of market by offering \gls{cdn} like services themselves. As a consequence of adopting \gls{sdn}, the \gls{cdn} can be easily and dynamically rearranged, the provider himself could interact with the system. Also the caching mechanism can be modified on demand as well as the content-to-cache assignment algorithm.


\subsection{Design}
\begin{figure}[t!]
\centering
\includegraphics[width=0.8\linewidth]{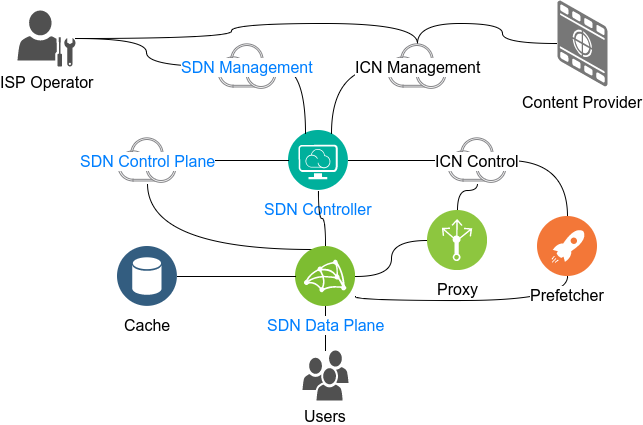}
\caption{Architectural overview showing administrative networks.}
\label{fig:clouds}
\end{figure}

The \gls{icnaas} vision considers five networks that interconnect the system actors as shown in Figure \ref{fig:clouds}. Apart from the typical \textit{SDN Control Plane} and \textit{SDN Data Plane}, as well as the \textit{SDN Management}, we define two new networks, the \textit{ICN Control} and the \textit{ICN Management}. The former is intended for communication between the \gls{icnaas} system, usually implemented as an \gls{sdn} application, with the \gls{icn} network elements in charge of offloading the \gls{sdn} controller from tasks related with \gls{icn} data transmission, while the later is employed by the Content Provider or the \gls{isp} operator to communicate with the system and arrange the \gls{icn} instances. This distinction comes from the functionality perspective and not from a real need of isolating such networks, the \textit{ICN and SDN Management} networks are probably simply the Internet while the \textit{ICN Control} and \textit{SDN Control Plane} can probably be the same collision domain, but the actors involved and the type of communication being carried on in each network is also a differentiating factor.

\begin{figure*}
\centering
\includegraphics[width=0.95\textwidth]{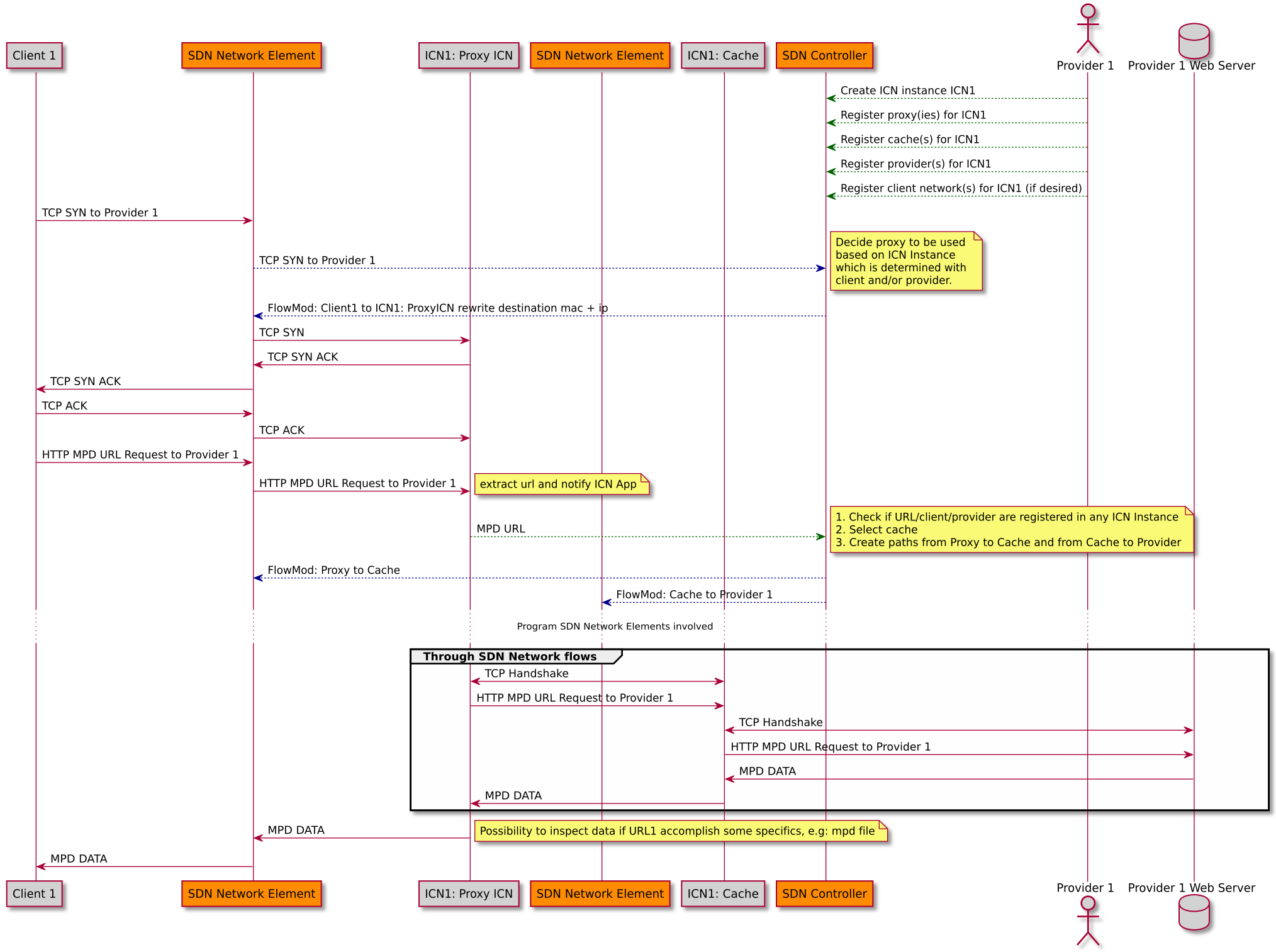}
\caption{Sequence Diagram of ICNaaS operations}
\label{fig:sequencediagram}
\end{figure*}

The interactions between the Content Provider and the \gls{icnaas} system to create, modify and remove proxies, caches and prefetchers (explained in Section \ref{sec:prefetch}) and the \gls{icn} instance themselves are provided through the \gls{rest} interface messages \#1-5 detailed in Table \ref{tab:Rest} and are represented in Figure \ref{fig:sequencediagram} (which shows the whole system working) with green arrows. 
The information necessary for each of these elements is the location, represented by the tuple (dpid,port) in \gls{openflow} and the link and network level information, meaning mac, ip and tcp port. These values are needed while steering the traffic to a certain element to be rewritten in the paths created with \gls{openflow} flows so that the operating systems running on the different elements of the network accept \textbf{transparently} the packets which otherwise were directed to other hosts of the network, therefore enabling the integration of any existing caching system in the architecture. Finally the Provider servers are to be registered as part of the icn, for these ones, standard ip routing is used and only the source network, the \gls{uri} pattern and host pattern are supplied for filtering which requests go to which icn instance. Source network serves as client filter and could be from any host to a precise host including a whole network mast, the \gls{uri} pattern filters the provider server to which it is making reference, take into account that any \gls{cdn} or \gls{dns} balancing mechanism existing outside of the \gls{sdn} network is still valid since the routing is based on the \gls{uri} and finally the host pattern, included in the \gls{icn} instantiation, employs the Host \gls{http} header to identify the requests related to this \gls{icn} instance.

\begin{table*}
\begin{center}
\begin{tabular}{|c|c|c|c|c|c|}
\hline
\# & Network  & Source   & Destination  & URL            & Parameters              \\ \hline\hline
1  & ICN Management & Provider & \gls{icnaas} & onos/icn/icn   & name, description, type \\ \hline
2  & ICN Management & Provider & \gls{icnaas} & onos/icn/proxy & name, description, mac, ip, proxy\_port, type,\\
&&&&& location (dpid,port), isProactive\\\hline
3  & ICN Management & Provider & \gls{icnaas} & onos/icn/prefetch & name, description, mac, ip, prefetch\_port, type, location (dpid,port)\\\hline
4  & ICN Management & Provider & \gls{icnaas} & onos/icn/cache & name, description, mac, ip, port, type, location (dpid,port)\\\hline
5  & ICN Management & Provider & \gls{icnaas} & onos/icn/provider   & instance, name, description, network, uripattern, hostpattern \\ \hline
6  & ICN Control    & Proxy    & \gls{icnaas} & onos/icn/proxyrequest & uri, hostname, smac, source ip, destination ip, protocol,\\
&&&&& source port, destination port\\\hline
7  & ICN Control    & \gls{icnaas} & Prefetcher & /prefetch & uri, server, port\\\hline
\end{tabular}
\end{center}
\caption{\gls{icnaas}'s REST interface.}
\label{tab:Rest}
\end{table*}

Once proxies and delivery networks have been setup through the public Northbound API , the \gls{sdn} application programs network devices to redirect \gls{http} requests targeted at a content provider towards the closest proxy, the redirection can be performed reactively once a TCP\_SYN message arrives to the controller or pro-actively. The proxy then uses the private Northbound API through what we denominated the \textit{ICN Control} network, detailed message \#6 in Table \ref{tab:Rest}, to notify the \gls{sdn} application about the requested resource (the \gls{url}, and the link, network and transport layer information). If such resource is not to be handled by any \gls{icn} instance, the controller (depending on the \gls{noc} policies) steers the traffic to the \gls{url} through the default gateway or discards host traffic. Otherwise, the application programs a bidirectional flow from the proxy to the most appropriate cache that holds such resource. Note that the \gls{icnaas} system is notified with the \gls{url} together with source address and port identifying each session to retrieve each \gls{url} independently. In case the resource is being requested for the first time, the application is responsible for choosing the most appropriate cache (according to the operator's policy) and programming the associated flows. Since this provokes a cache miss, the content must be downloaded from the origin server but it will be available for future requests.



In order to implement name-based content placement and retrieval, the \gls{sdn} application must inspect \gls{http} flows originated from consumers and targeted at providers within a delivery network. However, the application cannot find out the resource \gls{uri} until the \gls{tcp} three-way handshake has finished. This is problematic because the application must direct the flow to the appropriate cache or origin server since the first \gls{tcp} SYN segment. To overcome such issue, we have implemented a flexible \gls{http} proxy that performs delayed binding\cite{DelayedBindingKopparapu2002} (or \gls{tcp} splicing\cite{TcpSplicingBinder2015}) and provides our \gls{sdn} application with the name of the requested resource, details on the message sequence are shown in Figure \ref{fig:sequencediagram}.

The \gls{sdn} and in particular the \gls{openflow} protocol capabilities to rewrite message headers are used. In the case of the communication between the client and the proxy, the proxy receives the TCP\_SYN message with the destination mac, ip address and tcp port modified to those registered in the \gls{icnaas} system for that precise proxy, in previous approaches the receiver (let it be the proxy or the cache) needed to apply transparent proxy techniques and linux iptables solutions to accept packets directed to a different mac, different ip and sometimes even different port.  \gls{openflow} upon version 1.1.0\cite{Pfaff2011Openflow11} defines some \textit{Set-Field} type actions and in particular the \underline{optional} OFPAT\_SET\_DL\_DST for rewriting destination mac, OFPAT\_SET\_NW\_DST for rewriting the destination \gls{ip} address and the OFPAT\_SET\_TP\_DST which are used for forward connectivity flows, similarly for the backward connectivity flows for responses the optional OFPAT\_SET\_DL\_SRC, OFPAT\_SET\_NW\_SRC and OFPAT\_SET\_TP\_SRC are used to rewrite the source fields. Being optional means that it is not mandatory for a vendor to implement them and in some cases the actions are only available in software tables, meaning that they are not executed by the device hardware pipeline.

We envision the \gls{icnaas} as a three layer system: the \gls{icnaas} itself in charge of managing instances and steering traffic; a protocol specific layer aware of specifics about the service being provided with \gls{http} such as \gls{dash} or HTML; and finally the data layer with knowledge about the data itself, let it be \gls{svc} video or a simple JPEG. The caching location decision is performed between the last two layers and might be configurable by the Provider.



The presented system serves as the basis for enabling content aware \gls{http} prefetching systems.

\section{Prefetching \glspl{url}}
\label{sec:prefetch}
Thanks to the \gls{url} extraction mechanism and its provision to the \gls{icnaas} and by employing the correlation between requests in \gls{http}, the system can pro-actively request the content that is foreseen as to be requested as a consequence of the actual url. The simplest example is a web page in which the \textit{IMG} tags point to pictures that will presumably be downloaded just after the web page itself. The idea is to enhance the cache hit ratio therefore reducing the download time. To that end, we introduce the role of ’cache accelerator’ or 'prefetcher'. 

Some appliances might offer with the means to explicitly request for content caching, in that case the prefetcher might make use of that method and in some cases the \gls{icnaas} might itself contact the cache directly. In that case, the cache should be also connected to 'ICN Control' ('ICN Control' is shown in Figure \ref{fig:clouds}).

There is also the possibility to actually create fake HTTP requests that will trigger the caching mechanism transparently being vendor agnostic. 
The term fake here makes reference to the fact that the request is issued not by a customer but by the ICNaaS system itself predicting future requests and because requests are not performed fully but only the few first bytes are retrieved, avoiding unneeded network load. The cache accelerator could be perfectly implemented as part of the proxy reducing the number of trust relations of the ICNaaS application that sits on top of the ISP SDN controller that is a critical component of the network.

The introduction of the actual network state into the equation while steering the traffic to the caches has potential advantages such as the available bandwidth per link which for the case of video streaming might be used to force the users to a certain bit-rate version by dropping any request for higher bitrate versions.

%

\begin{figure*}
\centering
\includegraphics[width=\textwidth]{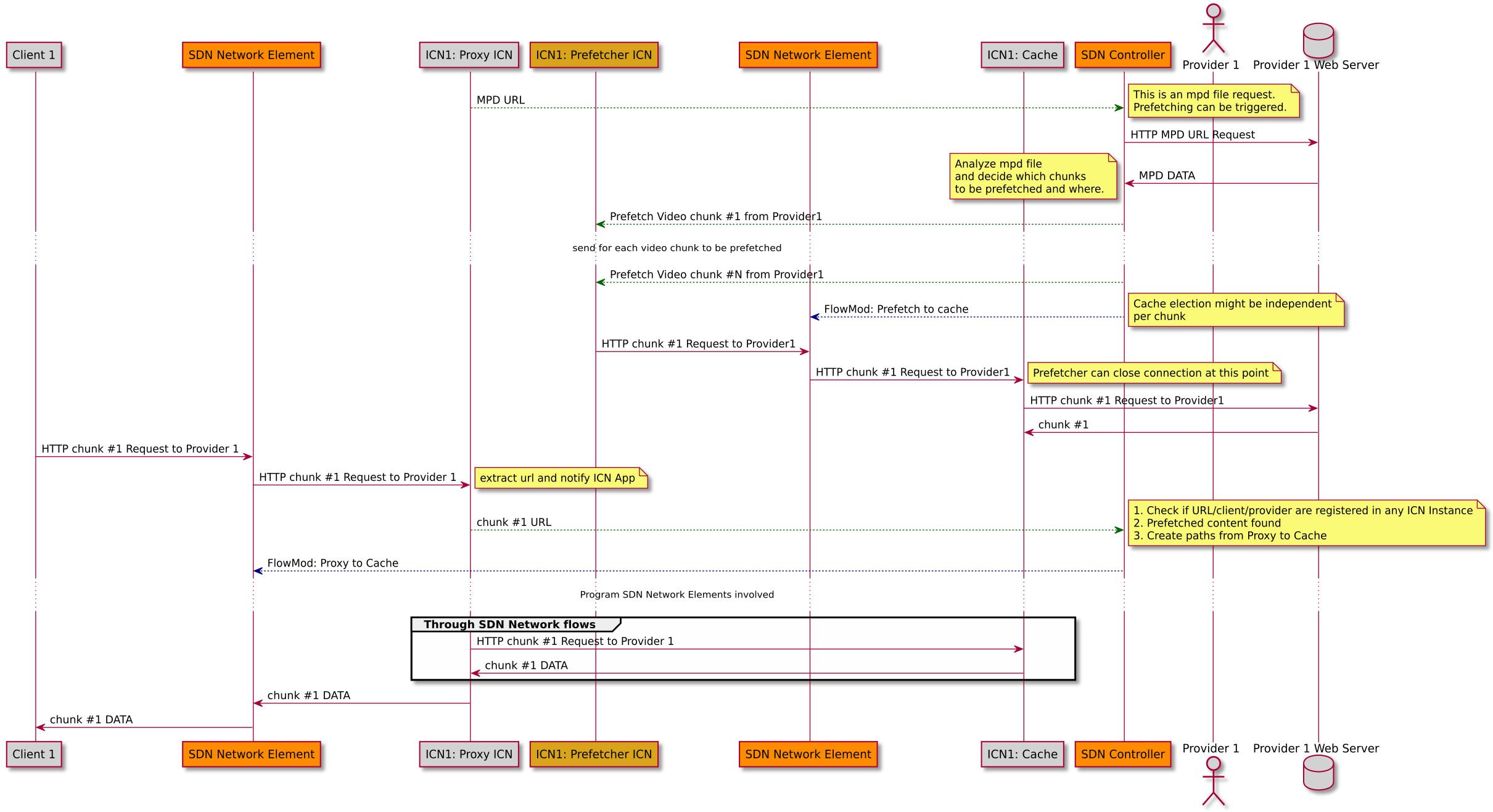}
\caption{Sequence Diagram for Prefetching mechanism}
\label{fig:sequencediagramprefetch}
\end{figure*}

As a consequence, the Prefetcher REST interface is defined as in Table \ref{tab:Rest} message \#7, to enable the communication of the \gls{icnaas} and the entity issuing the requests that will populate the cache prior to being requested. Note that in this case the \gls{icnaas} acts as a client and not as the service. This functionality could be implemented as part of the controller by means of OFPT\_PACKET\_IN and OFPT\_PACKET\_OUT messages which would in turn imply at least 5 \gls{tcp} messages (the 3-way handshake, the http request and the FIN). That approach would depend nevertheless on the caching entity behaviour when receiving the TCP FIN message, if it still continues downloading on the server side, it would be fruitful, if not, it would be a waste. On the other hand, this process would take these 5 messages per chunk which would rise linearly if there is dependency between chunks as is the case of \gls{svc} layers on \gls{dash} streams probably producing controller overload. The prefetcher, which could be collocated with the cache itself, avoiding network load, implements a full \gls{http} heap which means that issues requests identical to those issued by clients so that any caching system would be able to be used.




In the case of \gls{dash} video streaming the \gls{icnaas} detects in the protocol specific layer that an \gls{mpd} file has been requested. The system downloads the \gls{mpd} file in parallel to the client request (note that the \gls{mpd} data could also be supplied by the proxy but this keeps the proxy as simple as possible) and analyses it, storing the Representations \glspl{url} so that later each request can be identified with its Representation. When the client later requests for a certain \gls{url} the matching is performed and the corresponding Representation \glspl{url} are notified to the prefetcher for download.

If the data specific layer detects a scalable video codec, such as \gls{svc}, an extended matching is needed so that not only the Representation for that chunk is retrieved but also those on which the chunk layer depend.

We consider \gls{dash} with \gls{svc} a very interesting scenario for demonstrating the possibilities offered by metadata parsing for content prefetching, therefore we produced a proof of concept and evaluated it in the laboratory.

As a showcase of the possibilities offered by metadata parsing, we have implemented a rather simple \gls{svc} aware distributed caching system. The solution distributes the \gls{svc} layers over the caches available from the client to the network gateway uniformly, putting nearer to the client the lower scalability levels which have a higher probability to be requested. Thanks to the \gls{mpd} parsing process, the \gls{icnaas} knows exactly how many descriptions the video has and can assign each layer to a cache, depending on the number of registered caches. When a chunk is requested, the representation to which it pertains is matched knowing which other representations it depends on, with that information every possible url in the operation point is precomputed with a cache, so that next requests with related urls simply create the path to the corresponding cache. The algorithm implemented and evaluated in Section \ref{sec:results} is shown in Listing \ref{lst:distributedsvccachecode}.

\begin{lstlisting}[caption='Distributed SVC cache allocation algorithm.', label=lst:distributedsvccachecode, language=Java, basicstyle=\tiny]
Map<IMiddlebox, Integer> middleBoxesDistance =
    service.getMiddleBoxesDistance(caches.values(), sw);
Stream<Map.Entry<IMiddlebox, Integer>> sorted =
    middleBoxesDistance.entrySet().stream().sorted(new IntegerValueComparator());

// If it is an MPD just use the nearest cache
if (uri.endsWith(".mpd") || uri.endsWith("*.MPD")) {
    Map.Entry<IMiddlebox, Integer> iMiddleboxIntegerEntry =
        middleBoxesDistance.entrySet().stream()
        .min(new IntegerValueComparator()).orElse(null);
    if (iMiddleboxIntegerEntry != null)
        return (Cache)iMiddleboxIntegerEntry.getKey();
    return null;
}

ResourceHTTP resourceHTTP = new ResourceHTTP();
resourceHTTP.setFullurl(uri);
Optional<ResourceHTTP> resourceOpt =
    resources.values().parallelStream().filter(x -> {
        if (!x.getType().equals(ResourceHTTPDASH.DESCRIPTION))
            return false;
        ResourceHTTPDASH r = (ResourceHTTPDASH) x;
        return r.containsURL(uri);
    }).findFirst();

ResourceHTTPDASH rfull = (ResourceHTTPDASH)resourceOpt.orElse(null);
if (rfull == null) {
    log.error("There is no resource available for uri {}", uri);
    return null;
}

// Look for precomputed cache
ConcurrentHashMap<String, IMiddlebox> precomputedCaches =
    precomputedCachesXresourceXurl.getOrDefault(rfull.getFullurl(), null);
if (precomputedCaches != null) {
    log.info("Locating precomputed cache");
    IMiddlebox precomputedcache = precomputedCaches.getOrDefault(uri, null);
    if (precomputedcache != null) {
        return (Cache)precomputedcache;
    }
}

// Precompute all caches
log.info("Precomputing caches for {}", rfull.getFullurl());
ConcurrentHashMap<String, IMiddlebox> cacheXurl =
    new ConcurrentHashMap<>();
ConcurrentHashMap<String, Integer> fullUrlsRepresentationIds =
    rfull.getFullUrlsRepresentationIds();

Integer representationCount = rfull.getRepresentationCount();
Double dependencycacheratio =
    Math.ceil(representationCount / caches.size()); // Next integer
log.debug("DependenciesCache Ratio: {}", dependencycacheratio);
List<Integer> representationIds = rfull.getRepresentationIds();
List<IMiddlebox> orderedcachelist = middleBoxesDistance.entrySet().stream()
    .sorted(new IntegerValueComparator()).map(Map.Entry::getKey)
    .collect(Collectors.toList());

fullUrlsRepresentationIds.entrySet().parallelStream().forEach(x -> {
        int representationIdx = representationIds.indexOf(x.getValue());
        log.debug("Request representation index {} of {}",
                  representationIdx, representationCount);
        Double position =
            Math.floor(representationIdx / dependencycacheratio); // Previous integer
        log.debug("Position selected {}", position.intValue());
        IMiddlebox iMiddlebox = orderedcachelist.get(position.intValue());
        log.debug("Selected cache {}", iMiddlebox.getName());
        cacheXurl.put(x.getKey(), iMiddlebox);
    });

precomputedCachesXresourceXurl.put(rfull.getFullurl(), cacheXurl);

return (Cache)cacheXurl.getOrDefault(uri, null);
\end{lstlisting}

\section{Evaluating the Prefetching Mechanism}
\label{sec:results}

The architecture described in \cite{krishna2018user} and introduced in Section \ref{sec:icnaas} has been extended with the prefetching mechanism described in Section \ref{sec:prefetch} and has been evaluated in two different testbeds, one with real switches and real wiring and one completely virtualized employing mininet version 2.2.2 with \gls{ovs} version 2.9.2 for the network virtualization and \gls{lxc} on top of a dual socket 'Intel(R) Xeon(R) CPU E5-2603 v3 @ 1.60GHz'. This Section details the outcomes of the evaluation.

The \gls{icnaas} with the prefetching mechanism implemented as one \gls{icn} configurable alternative has been implemented on top of \gls{onos} controller. The proxy and the prefetcher have been implemented with Python 3 based on the \textit{tornado} library with a nginx proxy to allow request queueing. The caches are based on the well known Squid 3.5.12. For simplicity we deploy the prefetcher as a side entity of the proxy but it could be deployed anywhere into the network (even collocated with each cache).

The video employed for the evaluation is \url{http://concert.itec.aau.at/SVCDataset/dataset/mpd-temp/BBB-I-1080p.mpd} from the \textit{Institute of Information Technology, Alpen-Adria-Universität Klagenfurt, Austria}\cite{Kreuzberger2015} which contains 50 different \gls{svc} scalability levels and a duration of 10 minutes.

\begin{figure}
\centering
\includegraphics[width=\linewidth]{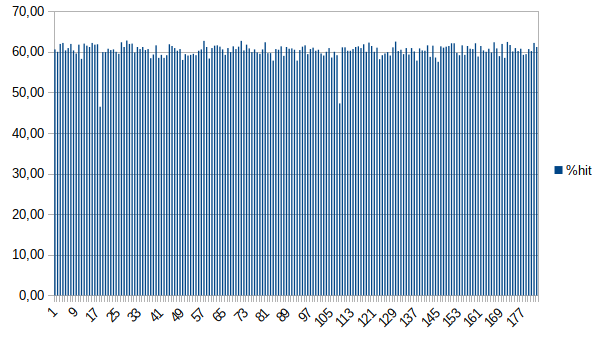}
\caption{Prefetch Cache Hit Ratio on Hardware switches}
\label{fig:hitratio}
\end{figure}

Our first evaluation has been carried out on an scenario with HPE ARUBA 2920 switches with software version WB.16.04.0008 configured to use OpenFlow 1.3, which in turn is the highest version available for these switches and the more stable version supported by \gls{onos}. In this evaluation dependencyId 18 is requested by the \gls{dash} client. Note that we deactivated the client side decision algorithm for the sake of focusing on the evaluation of the system and not in the effects of client decisions in caching systems as discussed above and introduced by Grandl et al.\cite{Grandl2013}.  In terms of cache hit ratio the prefetching mechanisms achieves in average $60.37\%$ (see Figure \ref{fig:hitratio}), the same scenario without prefetching mechanism achieves $98,66\%$ hit ratio on the second run, but it actually implies two full video streaming processes per run, 20 different runs were performed as base case. In average 2013 \gls{http} requests are performed to the cache of which 722 are hits that for each layer has a ratio of $L0 \rightarrow 57,87\%$, $L1 \rightarrow 76,26\%$, $L16 \rightarrow 51,50\%$ and $L17 \rightarrow 55,84\%$. In average the first hit is achieved 1,8 seconds after the \gls{mpd} file is retrieved while the precaching process is finished in mean 4 minutes and 43 seconds afterwards while the length of the video being streamed is 10 minutes.

\begin{figure}
	\centering
	\includegraphics[width=\linewidth]{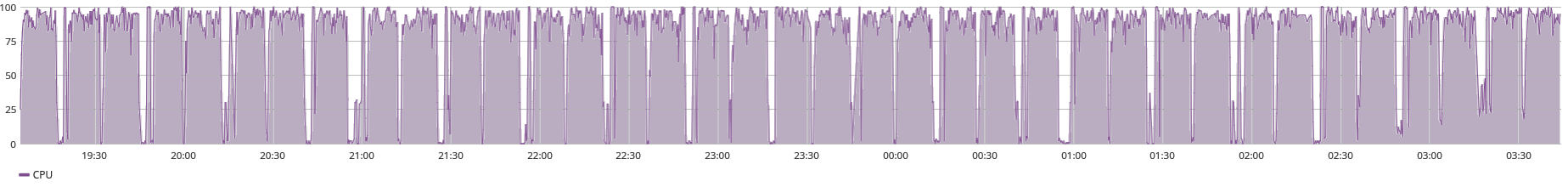}
	\caption{HP Hardware switch CPU overload}
	\label{fig:cpuoverload}
\end{figure}

The results obtained by the prefetching mechanism are far from the ones obtained in the base case. One of the problems found in our research is that HP switches do not support IP address rewriting in their pipelines. As a consequence the IP rewrite of each packet is performed in the switch main \gls{cpu} which is not intended for such a load. As can be seen in Figure \ref{fig:cpuoverload} the switch's \gls{cpu} is overwhelmed and therefore packet loss may be involved in the results, even \gls{onos} produces error logs regarding switch connectivity loss.

\begin{figure}
	\centering
	\includegraphics[width=\linewidth]{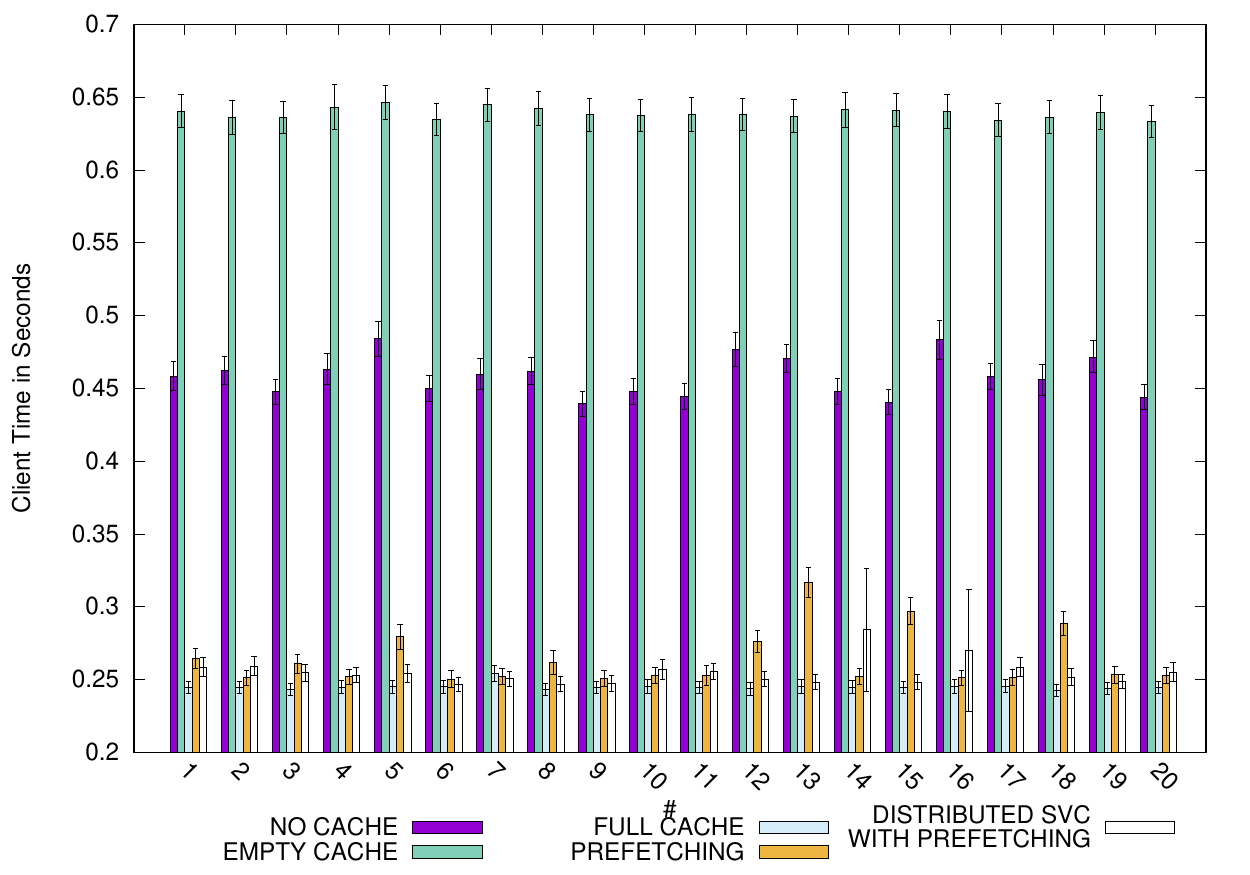}
	\caption{Client results}
	\label{fig:client results}
\end{figure}

In order to discard the hardware related problems, we migrated our testbed to mininet and connected the same software instances employed in the hardware scenario to the network clone. For this testbed a series of 20 run per case have been performed, being the cases analysed the corresponding to non cached retrieval as a reference, the empty cache case in which all the requests produce a cache miss event in the cache, just after that a full cache case is performed, again with empty caches the prefetching case is performed and finally the prefetching with distributed \gls{svc} cache allocation algorithm is evaluated. The results in terms of mean chunk download time are shown in Figure~\ref{fig:client results}. As expected, the direct connection case outperforms the results of accessing an empty cache but in turn get outperformed by the results obtained by any of the cached approaches which was the expected result. 4 out of 20 runs produce a slightly poorer performance for the prefetcher and distributed svc with prefetching case. We consider this slight differences a consequence of the prefetcher and proxy implementation that has connection retry on an exponential fashion. This was needed to overcome the \gls{onos} queuing approach to flow installation and could be avoided by "simply" employing flow addition listeners and notifying the two entities once the flows have been finally installed on the devices. We followed a more aggressive approach by notifying the entities once the controller accepts the request and making the entities retry connection attempts.

As part of the client chunk download time evaluation, the time employed by the proxy to notify the controller and get the answer notifying that channel to the cache for that precise chunk has been created has been analysed and is shown in Table \ref{tab:ctrl_time}. As can be seen the values are below 15 milliseconds which represents half of the time employed in accessing a cache with a HIT event.
\begin{table}
	\begin{tabular}{|c|c|c|c|c|}
		\hline
		CASE&MEAN CTRL&STDDEV&MIN&MAX\\\hline\hline
		CACHE EMPTY&0.0078&0.0004&0.0073&0.0086\\\hline
		CACHE FULL&0.0054&0.0003&0.0051&0.0064\\\hline
		PREFETCHER&0.0091&0.0018&0.0073&0.0143\\\hline
		DISTRIBUTED&0.0091&0.0013&0.0076&0.0122\\\hline
	\end{tabular} 
	\caption{Proxy interaction time with controller}
	\label{tab:ctrl_time}
\end{table}


As can be seen in Tables \ref{tab:cachefull_hitratio}, \ref{tab:prefetch_hitratio} and \ref{tab:dist_hitratio}, the achieved hit ratio is with the software switches which are not limited by the hardware pipeline limitations more in-line with the expected values around 90\% cache hit ratio. We have to take into account that most of the MISS events are cases in which the chunk had previously been requested but Squid has refused to cache it producing what in the logs is marked as 'TCP\_SWAPFAIL\_MISS', it is not the aim of this paper evaluate Squid since it is just one of the caching possibilities available in the market.

\begin{table}
	\centering
	\begin{tabular}{|c|c|c|c|c|}
		\hline 
		CASE&TOTAL&HIT&MISS&HIT RATIO\\\hline\hline
		01&1795&1742&53&97.0474\\\hline
		02&1795&1741&54&96.9916\\\hline
		03&1795&1741&54&96.9916\\\hline
		04&1795&1742&53&97.0474\\\hline
		05&1795&1740&55&96.9359\\\hline
		06&1795&1743&52&97.1031\\\hline
		07&1795&1742&53&97.0474\\\hline
		08&1795&1743&52&97.1031\\\hline
		09&1795&1740&55&96.9359\\\hline
		10&1795&1743&52&97.1031\\\hline
		11&1795&1742&53&97.0474\\\hline
		12&1795&1741&54&96.9916\\\hline
		13&1795&1743&52&97.1031\\\hline
		14&1795&1739&56&96.8802\\\hline
		15&1795&1742&53&97.0474\\\hline
		16&1795&1739&56&96.8802\\\hline
		17&1795&1741&54&96.9916\\\hline
		18&1795&1707&88&95.0975\\\hline
		19&1795&1743&52&97.1031\\\hline
		20&1795&1740&55&96.9359\\\hline
	\end{tabular}
	\caption{Cache hit ratio with cache FULL. Virtual switch scenario}
	\label{tab:cachefull_hitratio}
\end{table}

\begin{table}
	\centering
	\begin{tabular}{|c|c|c|c|c|}
		\hline 
		CASE&TOTAL&HIT&MISS&HIT RATIO\\\hline\hline
		01&1795&1704&91&94.9304\\\hline
		02&1795&1610&185&89.6936\\\hline
		03&1795&1699&96&94.6518\\\hline
		04&1795&1691&104&94.2061\\\hline
		05&1795&1531&264&85.2925\\\hline
		06&1795&1574&221&87.6880\\\hline
		07&1795&1701&94&94.7632\\\hline
		08&1795&1699&96&94.6518\\\hline
		09&1795&1699&96&94.6518\\\hline
		10&1795&1662&133&92.5905\\\hline
		11&1795&1700&95&94.7075\\\hline
		12&1795&1716&79&95.5989\\\hline
		13&1795&1693&102&94.3175\\\hline
		14&1795&1700&95&94.7075\\\hline
		15&1795&1685&110&93.8719\\\hline
		16&1795&1635&160&91.0864\\\hline
		17&1795&1710&85&95.2646\\\hline
		18&1795&1700&95&94.7075\\\hline
		19&1795&1631&164&90.8635\\\hline
		20&1795&1663&132&92.6462\\\hline
	\end{tabular}
	\caption{Cache hit ratio with Prefetching mechanism}
	\label{tab:prefetch_hitratio}
\end{table}

\begin{table}
	\centering
	\begin{tabular}{|c|c|c|c|c|c|c|c|}
		\hline 
		CASE&TOTAL&HIT&MISS&HIT&HIT&MISS&HIT\\
		&&C2&C2&RATIO&C1&C1&RATIO\\\hline\hline
		01&1795&585&13&97.8261&1132&65&94.5698\\\hline
		02&1795&587&11&98.1605&1131&66&94.4862\\\hline
		03&1795&581&17&97.1572&1126&71&94.0685\\\hline
		04&1795&582&16&97.3244&1149&48&95.9900\\\hline
		05&1795&583&15&97.4916&1142&55&95.4052\\\hline
		06&1795&582&16&97.3244&1135&62&94.8204\\\hline
		07&1795&581&16&97.3199&1131&66&94.4862\\\hline
		08&1795&582&16&97.3244&1130&67&94.4027\\\hline
		09&1795&581&17&97.1572&1138&59&95.0710\\\hline
		10&1795&588&10&98.3278&1118&79&93.4002\\\hline
		11&1795&585&13&97.8261&1142&55&95.4052\\\hline
		12&1795&588&10&98.3278&1116&81&93.2331\\\hline
		13&1795&579&19&96.8227&1119&78&93.4837\\\hline
		14&1795&585&13&97.8261&1130&67&94.4027\\\hline
		15&1795&586&12&97.9933&1137&60&94.9875\\\hline
		16&1795&585&13&97.8261&1129&68&94.3191\\\hline
		17&1795&588&9&98.4925&1136&61&94.9039\\\hline
		18&1795&581&17&97.1572&1131&66&94.4862\\\hline
		19&1795&582&16&97.3244&1118&79&93.4002\\\hline
		20&1795&580&18&96.9900&1129&68&94.3191\\\hline
	\end{tabular}
	\label{tab:dist_hitratio}
	\caption{Cache hit ratio with Distributed Prefetching mechanism}
\end{table}

\begin{figure}
	\centering
	\includegraphics[width=\linewidth]{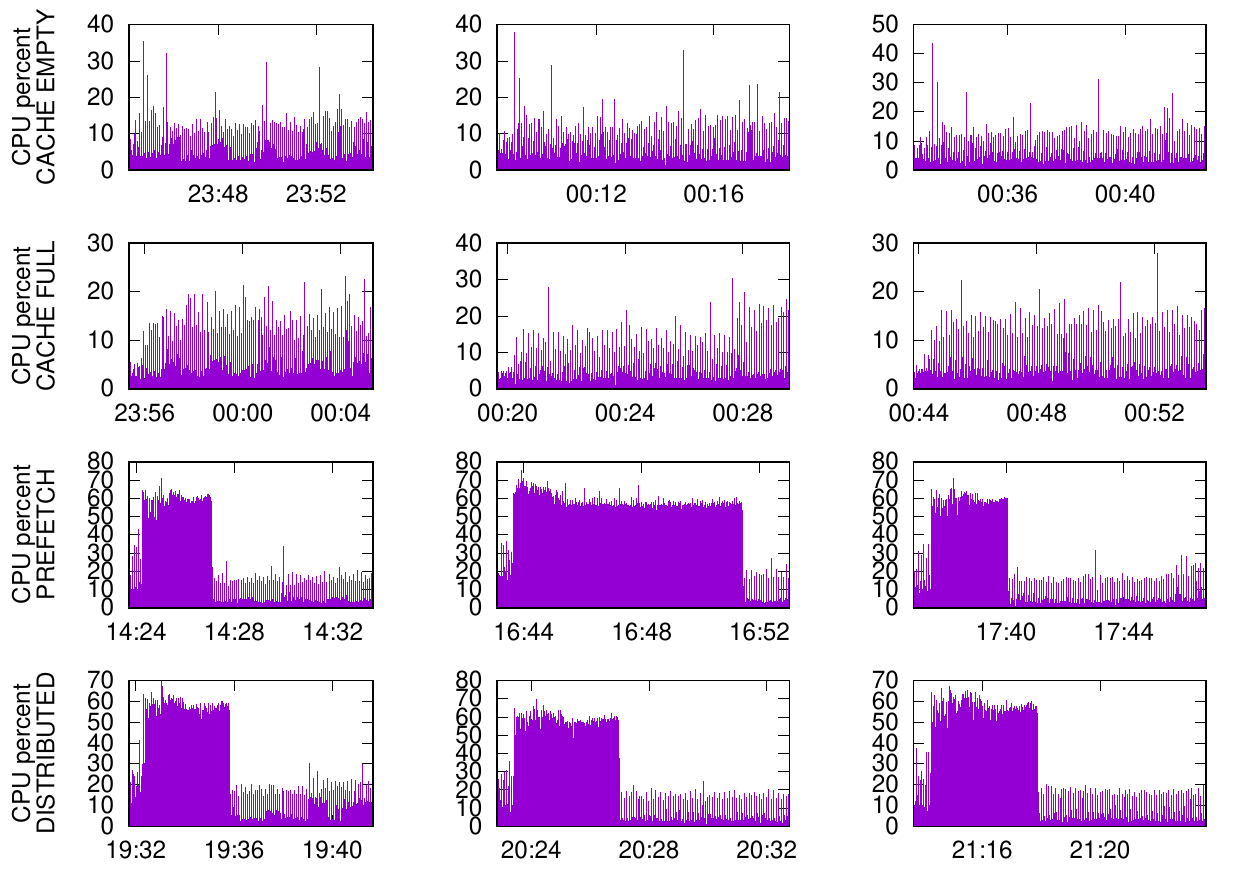}
	\caption{Controller CPU usage}
	\label{fig:controllercpu}
\end{figure}

\begin{figure}
	\centering
\includegraphics[width=\linewidth]{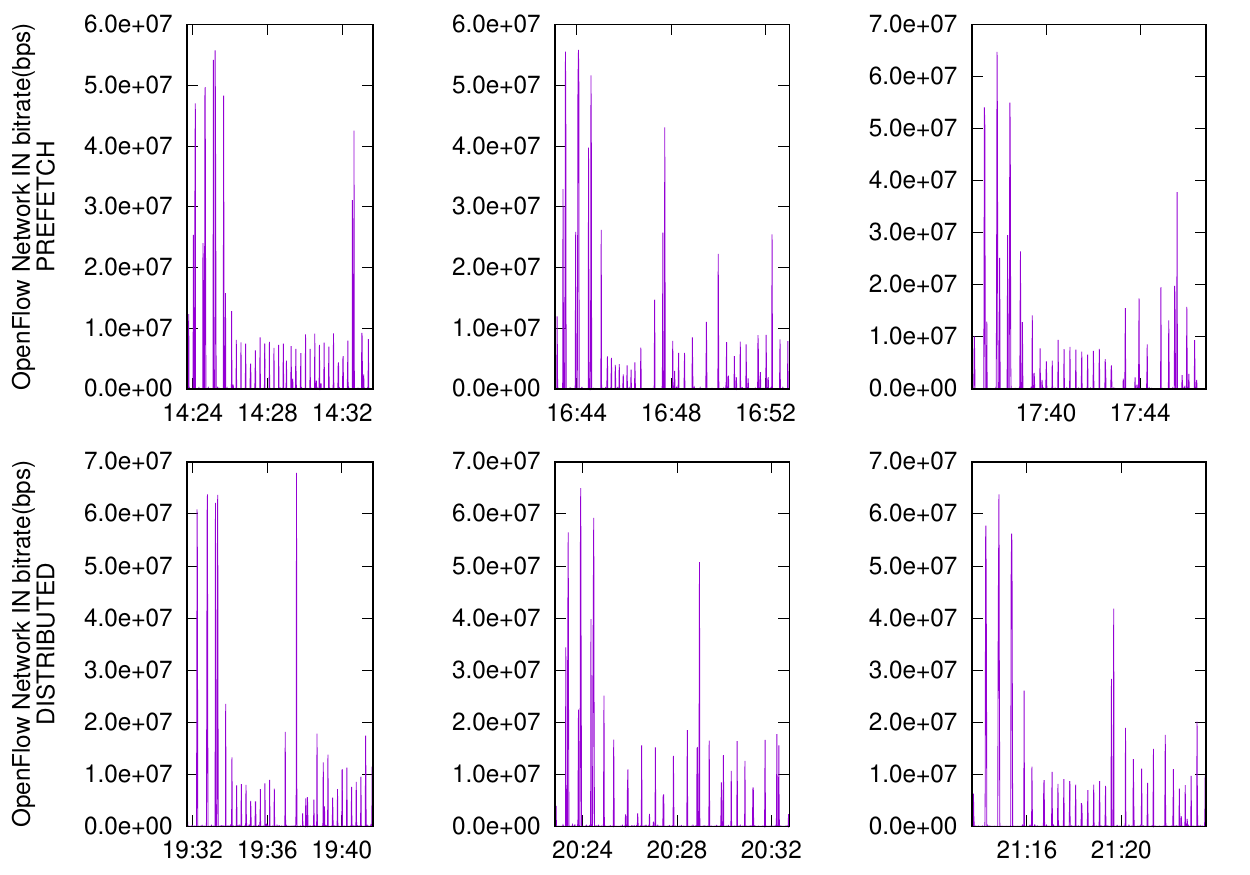}
\caption{Controller network usage}
\label{fig:controllernet}
\end{figure}

Even though the migration to mininet has produced better results for the hit ratio, it is still far from the $98,66\%$ cache hit ratio. The controller plays an active role in our prefetching solution which forces to maintain information about the data sources such as the mpd parsed information. Another possible approach would be to delegate fully the prefetching mechanism to another entity which would inform the controller via REST API. This solution would also reduce the cpu consumption of the controller but would reduce the possibilities for future caching decision taking algorithms. The \gls{cpu} and network load is shown in Figures \ref{fig:controllercpu} and \ref{fig:controllernet} for three random runs of each scenario. As can be seen there is a \gls{cpu} spike at the beginning of the prefetching enabled cases as a consequence of \gls{mpd} analysis and calculation of caches. The network usage on the other hand stays with spikes below 60Mbps which includes \gls{openflow} traffic as well as prefetcher and proxy signaling. Since the network load is not extended in long periods of time, roughly a few seconds, we don't foresee it as a problem.

The prefetching with distributed \gls{svc} cache allocation algorithm evaluation results have already been exposed in Table \ref{tab:dist_hitratio} in which two caches have been employed. As can be seen, C1 (standing for cache 1, the nearest from the client) receives more requests than C2. This is like that because the representation requested is 33 from 50 where dependencies for the latter are "49 48 34 33 32 18 17 16 2 1 0". Following the algorithm described in Listing \ref{lst:distributedsvccachecode} the Ids are distributed to C2 for those over 18 and those below 32 to C1, since the experiment is requesting Id 33, only 2 layers are stored in C2 and 6 layers are stored in C1.


\section{Conclusions and Future Work}
\label{sec:conclusions}
We presented a system to implement \gls{icnaas} with end-to-end \gls{http} communication transparent to the end points involved and capable of integrating legacy servers and caching systems.

On top of that system, we have envisioned a way to predict which content is going to be retrieved thanks to the correlation usually present in \gls{http} solutions and have integrated it into the \gls{icnaas} system to finally evaluate it. The evaluation has shown the feasibility of the proposal, the possibilities of the prefetching mechanism thanks to metadata parsing and has show one specific instantiation with the Distributed SVC caching allocation mechanism opening a new field of research. Nevertheless, the high \gls{cpu} usage has to be taken into account for future studies by probably off-loading the controller to a side entity.

Another field of research opened by metadata handling in the \gls{sdn} controller is the security of the controller which, even if it is already an active field, now is extended by the possibility to attack the controller by malicious metadata files, such as a intentionally deployed \gls{mpd}.

Two are the next steps in our research line. First is to investigate the caching distribution algorithms mentioned to provide alternatives for different goals, such as reduce the zapping time or save bandwidth between the leafs and the root of the network. Second is the inclusion of this \gls{icnaas} in the MANO architecture and how the later could take care of the deployment of the proxies, prefetchers and if needed the caches and register them for the Provider.

\section{Acknowledgment}
This paper has been funded by the H2020 EU project ANASTACIA project, Grant Agreement N 731558 and by the GN4-2 project under Grant Agreement No. 731122.

\bibliographystyle{plain}
\bibliography{references}
\end{document}